\documentclass[rmp,10pt]{revtex4}
\usepackage[utf8]{inputenc}
\usepackage{verbatim}
\usepackage[sort&compress]{natbib}
\setcitestyle{numbers}
\usepackage{mathtools}
\usepackage{amsmath}
\usepackage{amsfonts}
\usepackage{amssymb}
\usepackage{epsfig}
\usepackage{wrapfig}
\usepackage{changepage}
\usepackage{float}
\usepackage{wrapfig}

\newcommand{\beq}{\begin{equation}}
\newcommand{\eeq}{\end{equation}}

\begin{document}
\title{A general efficiency relation for molecular machines}

\author{Milo M. Lin}
\affiliation{Green Center for Systems Biology, Department of Bioinformatics, Department of Biophysics, and the Center for Alzheimer's and Neurodegenerative Diseases, University of Texas Southwestern Medical Center, Dallas, TX 75390 \\
Milo.Lin@UTSouthwestern.edu}

\begin{abstract}
Living systems efficiently use chemical fuel to do work, process information, and assemble patterns despite thermal noise. Whether high efficiency arises from general principles or specific fine-tuning is unknown. Here, applying a recent mapping from nonequilibrium systems to battery-resistor circuits, I derive an analytic expression for the efficiency of any dissipative molecular machine driven by one or a series of chemical potential differences. This expression disentangles the chemical potential from the machine's details, whose effect on the efficiency is fully specified by a constant called the load resistance. The efficiency passes through a switch-like inflection point if the balance between chemical potential and load resistance exceeds thermal noise. Therefore, dissipative chemical engines qualitatively differ from heat engines, which lack threshold behavior. This explains all-or-none dynein stepping with increasing ATP concentration observed in single-molecule experiments. These results indicate that biomolecular energy transduction is efficient not because of idosyncratic optimization of the biomolecules themselves, but rather because the concentration of chemical fuel is kept above a threshold level within cells.     
\end{abstract}

\maketitle



Energy is a limiting resource for life. Therefore, molecular machines that harness differences in chemical potential $\Delta \mu$ to perform tasks within the cell must function as highly efficient chemical engines. Despite thermal stochasticity, the measured efficiencies of biomolecular chemical engines are consistently in the 60-to-90 percent range \cite{kinesin_efficiency,f1atpase_oster, metabolism_efficiency}, well above that of typical human-designed systems such as heat engines. Two centuries ago, Carnot showed that the maximum efficiency of heat engines is \cite{carnot}:
\begin{equation}
 \eta_{\mathrm{heat}} \leq {{\Delta T}\over{\Delta T+T_{\mathrm{0}}}}.
 \end{equation} 
This relation depends on two independent parameters: the temperature differential driving the engine $\Delta T$ and the exhaust temperature $T_0$. Eq. 1 provides a universal constraint for all heat engines regardless of design specifics, and initiated the field of thermodynamics. Finding a general relation constraining the behavior of chemical engines would provide a unifying framework for biomolecular function and evolution, and for engineering nanodevices for physiological conditions.
\par
It is important to distinguish two broad classes of chemical engines in biology, as they should have different formulations of efficiency as well as design constraints. The first is the class of energy transduction machines, such as ion pumps, that convert energy from one chemical reservoir to another. In these cases, it is clear that any energy dissipated by the machine is wasted and so should be minimized in order to maximize the efficiency of transferring energy between the reservoirs, which is achieved in the "tight-coupling" limit \cite{effberg}. 
\par
This work is primarily concerned with the second class of chemical engines whose purpose \textit{is} the dissipation of the input energy. This class of dissipative machines serve a wide range of functions, including mechanical transport against viscous drag, and maintaining patterns of protein assembly or signaling configurations that are inaccessible at equilibrium (Fig. 1A). They are therefore principal agents of information processing and structural reorganization within the cell. Prigogine called these configurations "dissipative structures" because input energy is constantly dissipated to the environment as heat to maintain these patterns, which are far from equilibrium even under steady state conditions \cite{prigogine2}. Chemical engines can approach perfect efficiency if the throughput and power approach zero \cite{effberg,effprost}. Therefore, unlike Carnot's relation Eq.1, a useful general relation for the efficiency of chemical engines must depend on at least one system-dependent collective variable that reflects the constraint on throughput in the high efficiency limit, and reveal if there are regimes of the collective variable and $\Delta \mu$ for which the machine is both fast and efficient. Therefore, such a relation could also reveal if achieving high performance in both metrics requires evolutionary fine-tuning, or is a general property. We currently lack mechanistic insight into how chemical engine efficiency is controlled by tunable parameters. And it is unclear if the multitude of system-specific parameters could be encapsulated by a single collective variable that is well defined for all systems, which is necessary for a clear unified understanding.
\par
Existing results on dissipative chemical engines, based on generalized fluctuation-dissipation relations \cite{barato, todd}, upper-bound the efficiency relative to observed fluctuations \cite{seifert_2011, max_eff_universal}, although the bound may not be tight (achievable) far from equilibrium. Existing work does provide efficiency constraints under limiting conditions. For example, weakly driven engines operating at maximum power must be 50\% efficient \cite{eff_increase, broeck, lindenberg}. However, this bound only holds near equilibrium, and does not apply to the strongly-driven conditions relevant for biology. Numerical simulations of simple chemical engines indicate that efficiency is increased when driven farther from equilibrium by increasing $\Delta \mu$, similar to the effect of increasing $\Delta T$ in heat engines \cite{eff_increase}. This is consistent with a model in which molecular motors can become less wasteful if driven by higher ATP concentration \cite{thirumalai}, and suggest a general link between large $\Delta \mu$ and high efficiency. 

Here, using a recent generalization of the Boltzmann distribution to nonequilibrium systems\cite{circuit}, I derive a general equation for the maximum efficiency of dissipative chemical engines. This relation depends on two independent tunable variables: $\Delta \mu$ and the relative load resistance. The load resistance is a collective variable condensing the engine details, and is a constant unaffected by how strongly the engine is driven (i.e. $\Delta \mu$). The equation shows generally that biomolecular processes are intrinsically efficient because $\Delta \mu$ is much larger than the thermal energy $kT$ \textit{in vivo}. However, unlike heat engines, chemical engines have an efficiency inflection point at a threshold $\Delta \mu/kT$ that depends on the load resistance. This switch-like transition explains single-molecule \textit{in vitro} measurements of dynein molecular motor stepping. This result also leads to a general tradeoff relation between energy efficiency and time efficiency. More broadly, the existence of a general efficiency inflection point provides a candidate thermodynamic necessary condition for life.
 \newline
 \newline
 \textbf{Circuit mapping.} Chemical engines within the cell (circular arrows in Fig. 1A) consume energy to form patterns of activated signaling molecules, protein self-assembly, or directed movement, that would be vanishingly rare at equilibrium. Such processes can be modeled as nonequilibrium Markov chains, but obtaining analytical insights into such systems has mostly been intractable \cite{hill,schnakenberg,zia}. This work exploits a recently found mapping from the dynamical network of any Markovian system to an electronic circuit \cite{circuit} (See, for example, Fig. 1C,D). Consequently, a system is decomposed into a passive equilibrium portion ("resistors") that is driven by chemical potential differences $\Delta \mu$ ("batteries"). The resistance between two neighboring states $m$ and $n$ is: $R_{mn} \equiv {{\mathrm{e}^{\beta G_m}} \over {k_{mn}}}$, where $\beta = {{1}\over{kT}}$, $G_m$ is the free energy of state $m$ at equilibrium (i.e. in the absence of driving), and $k_{mn}$ is the equilibrium rate constant of transitioning from state $m$ to $n$. If a transition is directly driven by $\Delta \mu$, this transition is mapped to a battery with a probability potential drop $\mathcal{E}_{mn} \equiv (\mathrm{e}^{\beta \Delta \mu}-1) \mathrm{e}^{\beta G_m}P_{m}$ (red arrow in Fig. 1B). Using this mapping, probability flow amongst the states of any system obeys Ohm's law \cite{circuit}:
$P_{j} \mathrm{e}^{\beta G_j} -P_{i} \mathrm{e}^{\beta G_i} = {\sum\limits_{m=i}^{n=j}(\mathcal{E}_{mn}-R_{mn}I_{mn})}$ \cite{circuit}, where the summation is along any trajectory (parameterized by neighboring states $m,n$) connecting states $i$ and $j$. $I_{mn}$ is the probability current (net flux) from state $m$ to state $n$. Note that $\mathcal{E}_{mn}$ and $I_{mn}$ are zero for a system at equilibrium, and Ohm's law reduces to the Boltzmann distribution $P_{j} = P_{i} \mathrm{e}^{\beta (G_i-G_j)}$ in this case, as expected.
 In this mapping, multiple interconnected dynamical transitions can be systematically combined into a single resistor, called the Thevenin equivalent resistance \cite{Thevenin2}, that captures their collective effect on the probability flux at steady state.
 \begin{figure}
    \centerline{\includegraphics[width=\textwidth]{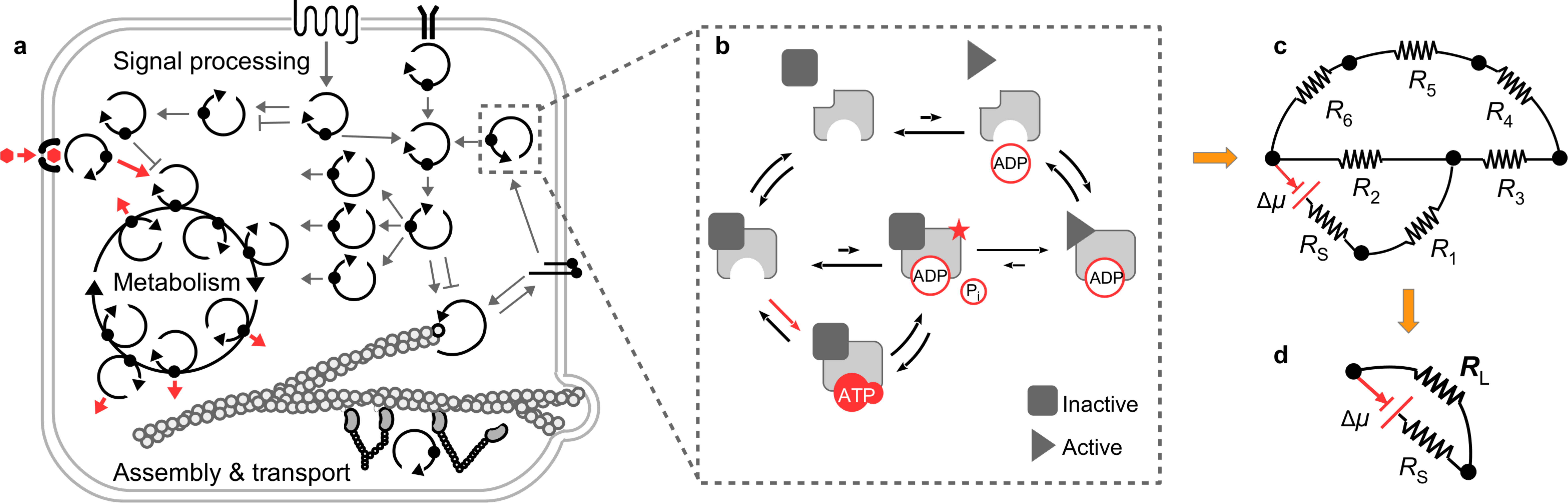}}
    \caption{\textsf{\textbf{Mapping nonequilibrium cycles to battery-resistor circuits.} Cellular function and homeostasis are maintained by cyclic molecular engines that harness chemical energy to transform and organize matter (simplified schematic of a cell in A; energy flow in red). The circuit mapping transforms a dynamical system, for example an enzymatic engine that activates a signaling molecule using energy released from hydrolyzing ATP into ADP and phosphate (B; star denotes activated enzyme complex), into an electronic circuit consisting of batteries and resistors that obey Ohm's law (C). Subsequently, the circuit elements can be systematically simplified to describe coarse-grained probabilities and currents without loss of accuracy (D).}}
    \label{fig1}
\end{figure}
 For example, for the signaling protein in Fig.1B, molecular transitions corresponding to $R_1,..,R_6$ (Fig. 1C) can be combined as series and parallel resistors into the equivalent resistance $R_{\mathrm{L}} = R_1 + ({1\over{R_2}}+{1\over{R_3+R_4+R_5+R_6}})^{-1}$ (Fig. 1D). Resistor coarse-graining, combined with the broad set of circuit theorems, allows a systematic approach to simplify systems of otherwise intractable complexity \cite{circuit}. This approach is most powerful for systems, such as biomolecular machines, for which the driven transitions are sparse.  
\newline
\newline
\textbf{Defining the efficiency of dissipative chemical engines.} 
Dissipative machines within the cell are usually composed of proteins that occupy one of many possible conformation and interaction states (Fig. 1B). Each machine harnesses a source of chemical energy, often by binding excess ATP and coupling its hydrolysis into ADP and phosphate to drive the target reaction. Consider the circuit representation of a machine driven by the source transition $\mathrm{S}$ (i.e. battery) with $\Delta \mu= kT\ln{\big[1+{{\alpha_{\mathrm{S}}}\over{k_{\mathrm{S}}}}\big]}$, where $k_{\mathrm{S}}$ is the apparent rate constant of the source transition in the absence of driving (e.g. no excess ATP at equilibrium), and $\alpha_{\mathrm{S}}$ is the enhancement in the rate constant due to driving. If the energy source is ATP, the source transition is enhancement of the rate of converting unbound phosphate and ADP into ATP bound to the protein (red arrow in Fig. 1B). At ATP concentrations well below binding saturation, $\alpha_{\mathrm{S}}$ is proportional to ATP concentration in excess of the equilibrium concentration. The resistance of the source transition (i.e. battery) is denoted $R_{\mathrm{S}}$ (Fig. 2A). A machine may contain multiple source transitions. Although the circuit framework is compatible with arbitrary arrangements of batteries, the analytic results in this work are restricted to those machines with a single source transition or multiple source transitions in series. Nevertheless, this class of machines includes a large fraction of cellular processes including signaling, self-assembly and sorting, and mechanical transport.
\par
The source transition breaks detailed balance for the machine by driving net flux of downstream transitions between states, and this flux may take numerous alternate pathways through state space. All transitions aside from the source transition can be combined into a single Thevenin equivalent load resistor $R_{\mathrm{L}}$ (Fig. 2A). Often, the biologically useful output of the engine is to enhance a particular state within the load. 
\par
With this formulation, we can now define the maximum efficiency of a dissipative machine as the dissipation rate of the load divided by the total dissipation rate. This definition corresponds to existing formulations of efficiency as special cases. For example, in the case that the machine is a molecular motor transporting a cargo against viscous drag, the load dissipation coincides with the maximum mechanical work that can be done by the motor per cycle. In this case, the efficiency is equivalent to the Stokes efficiency, which is the average viscous drag times the average velocity divided by the total dissipation rate \cite{Stokes_eff}. More broadly, this definition also extends the concept of efficiency to most processes in the cell, such as signaling and pattern formation, in which the useful dissipation of energy does not correspond to work along a simple reaction coordinate.
\newline
\newline
\textbf{A general efficiency relation for dissipative chemical engines.} We first obtain the total dissipation rate $\sigma T$, where $\sigma$ is the entropy production rate. The nonequilibrium fluctuation theorems \cite{evans, Jarzynski, Crooks} relate dissipation rate to transition probabilities: $\sigma T = kT\sum_{ij} I_{ij} \ln {\Bigg[ {{P_i(k_{ij}+\alpha_{ij})}\over{P_j k_{ji}}} \Bigg]}$ \cite{schnakenberg}. Because dissipation is summed over all possible dynamical transitions of a system, it may appear that knowing an engine's details is required to calculate the efficiency. Instead, I show here that $R_{\mathrm{L}}$ is the only information about the load needed to calculate the dissipation, and therefore the efficiency. First, using Tellegen's Circuit Theorem \cite{tellegen}, the total steady-state dissipation rate is (See Supp. Information):
\begin{equation}
 \sigma T= \sum_{ij} I_{ij} \Delta\mu_{ij}.
 \end{equation} 
In this reformulation, the total dissipation rate is a function of only the currents of the directly driven transitions (for which $\alpha_{ij} \neq 0$), even though energy is dissipated at all transitions. Eq. 2 is especially useful if driven transitions are sparse.  \par
\begin{figure}
    \centerline{\includegraphics[width=\textwidth]{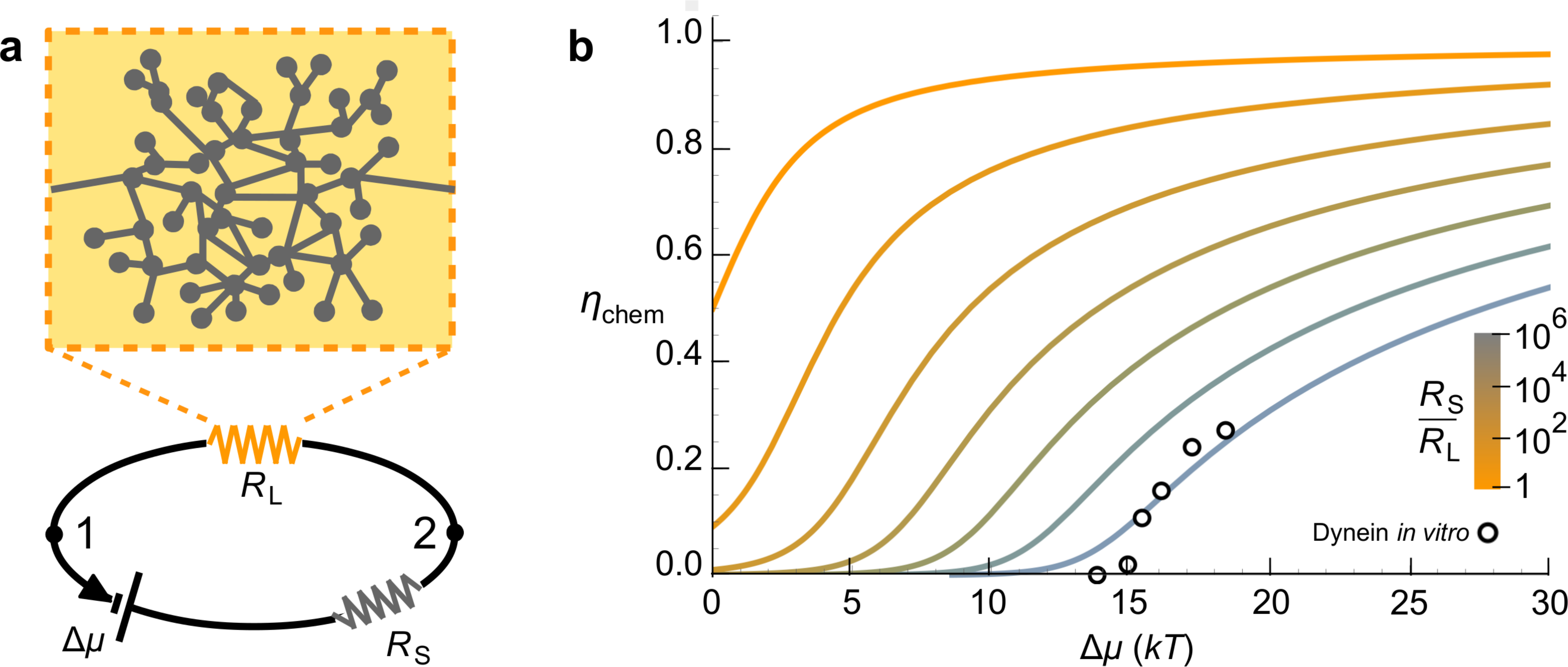}}
    \caption{\textsf{\textbf{Maximum efficiency of a dissipative machine.} The circuit for a dissipative machine is divided into the source resistance being directly driven by the chemical potential difference, and the load resistance comprising the target task being performed by the engine (orange inset), which may involve dynamical transitions between many states of the system (A). Eq. 3 predicts the maximum achievable efficiency, which is plotted as a function of $\Delta \mu/kT$ for different values of the relative load resistance (gray to orange lines; adjacent lines differ in $R_{\mathrm{S}}/R_{\mathrm{L}}$ by one order of magnitude) (B). The experimentally measured efficiency of dynein molecular motor as a function of ATP concentration \textit{in vitro} is adapted from Ref. 28 (open circles).}}
    \label{fig2}
\end{figure}

 If there is a single driven transition, the steady-state load dissipation rate is (Supp. Information): $\sigma_{\mathrm{L}}T= kTI_{\mathrm{S}} \ln{\bigg[{{R_{\mathrm{S}}+R_{\mathrm{L}}e^{\Delta \mu/kT}}\over{R_{\mathrm{S}}+R_{\mathrm{L}}}}  \bigg]}$. The difference between $\sigma T$ and $\sigma_{\mathrm{L}}T$ is the dissipation of the source (battery), which corresponds to energy that is not accessible to the load. The efficiency is upper-bounded because of the dissipation of the source, which corresponds to the energy lost whenever the engine dynamics happens to reverse the source transition, for example, if bound ATP is transformed into unbound ADP and phosphate without coupling to any downstream load reactions. The maximum steady state efficiency is: $\eta_{\mathrm{chem}} \leq \sigma_{\mathrm{L}}/ \sigma$. This yields the main result of this work, a general relation governing the maximum efficiency of dissipative chemical engines:

\
\begin{equation}
\eta_{\mathrm{chem}} \leq {{kT}\over{\Delta \mu}} \ln{\Bigg[{{{{R_{\mathrm{S}}}\over{R_{\mathrm{L}}}}+e^{{{\Delta \mu} \over{kT}}}}\over{{{R_{\mathrm{S}}}\over{R_{\mathrm{L}}}}+1}}  \Bigg]}.
\end{equation}
The efficiency depends on the chemical potential difference relative to $kT$, which quantifies how the efficiency is limited by  thermal fluctuations. The functional form of Eq. 3 remains valid in the case that the engine is cumulatively powered by multiple driven transitions (e.g. sequential hydrolysis of multiple ATP molecules); in this general case, $\Delta \mu$ is replaced by the sum of $\Delta \mu$'s and $R_{\mathrm{S}}$ is replaced by the weighted sum of all driven resistors (See Supp. Information). An engine will approach the maximum efficiency set by Eq. 3 if the most dissipative reactions within the load are useful. In practice, the load may include nonproductive futile cycles which dissipate energy subsequent to hydrolysis. 
\par
Some features of Eq. 3 are reminiscent of Carnot's heat engine law Eq. 1. (i) As with Eq. 1, Eq. 3 is a tight upper bound achievable if there are no futile processes in the load. (ii) Although Eq. 3 is a function of the constituent parameters, it is not dependent on knowing details of the engine. (iii) Eq. 3 predicts that the maximum efficiency of a dissipative machine increases with increasing $\Delta \mu/kT$, similar to how heat engine efficiency increases for larger $\Delta T$. This explains the observed increase in isothermal ratchet and motor efficiency when strongly driven \cite{efficiency, motor}. 
Despite these similarities, Eq. 1 (heat engines) and Eq. 3 (dissipative chemical engines) differ in fundamental ways. The efficiency of heat engines is a monotonic concave function of $\Delta T$, with no well-defined separation between inefficient engines near equilibrium and efficient engines far from equilibrium. In contrast, Eq. 3 shows that, in the functionally relevant processive regime ($R_{\mathrm{S}} > R_{\mathrm{L}}$; See Supp. Information), the efficiency of a dissipative chemical engine is a sigmoidal function of $\Delta \mu$. The inflection point of the sigmoid corresponds to a threshold $\Delta \mu_{\mathrm{thresh}}$:
\begin{equation}
\Delta \mu_\mathrm{thresh} \approx kT\ln\bigg[{{R_{\mathrm{S}}}\over{R_{\mathrm{L}}}}\bigg].
\end{equation}
\begin{wrapfigure}[20]{r}{0.48\textwidth}
  \begin{center}
    \includegraphics[width=0.47\textwidth]{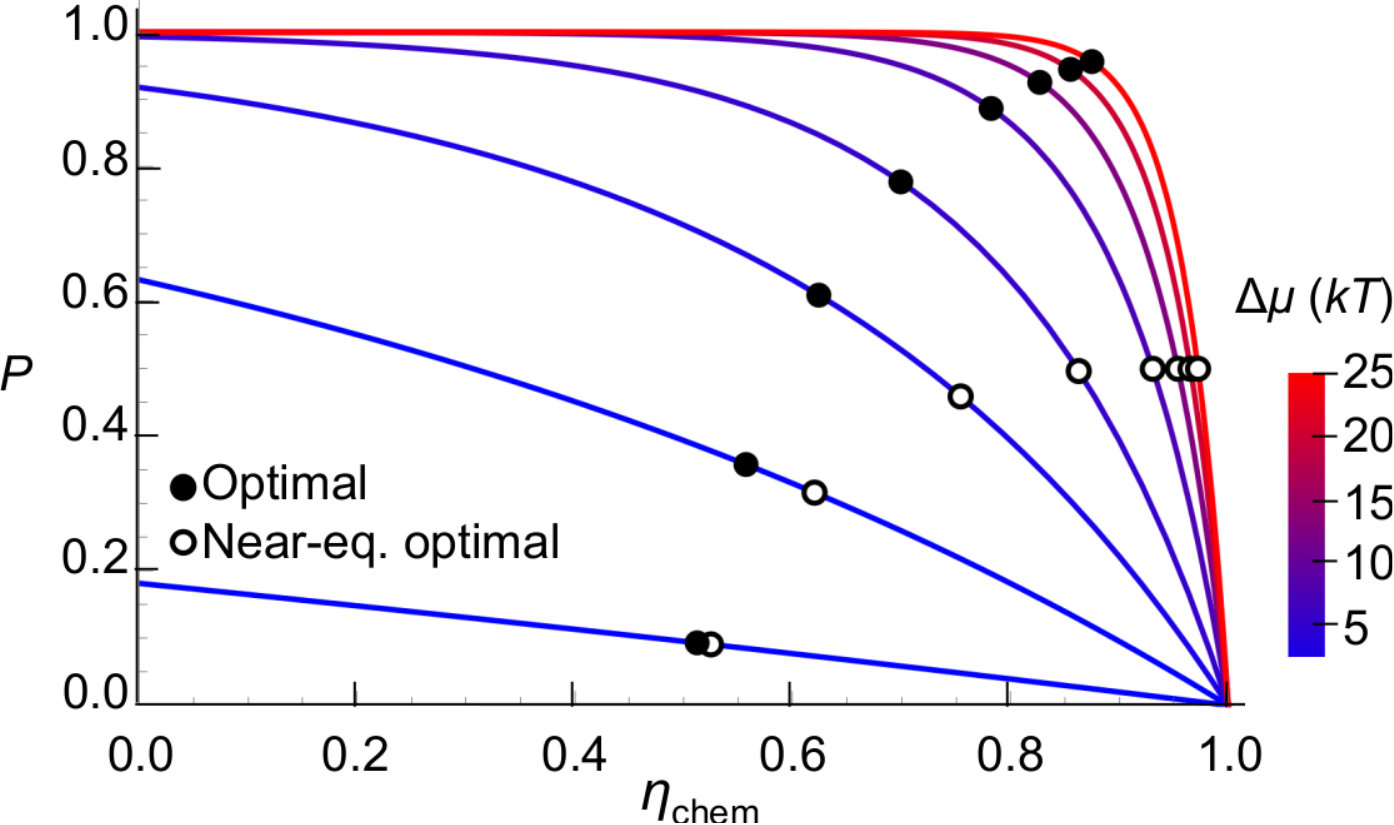}
  \end{center}
  \caption{\textsf{\textbf{Tradeoff between time and energy efficiency.} Eq. 5 is plotted for different values of $\Delta \mu$. Time efficiency, quantified by the processivity $P$, also varies from 0 to 1. Black filled circles denote optimal engines that maximize the sum of both efficiencies. White circles denote engines for which $R_{\mathrm{L}}=R_{\mathrm{S}}$}.}
\end{wrapfigure}
Below this threshold, all dissipative machines are inefficient. The efficiency only responds to increasing $\Delta \mu$ if $\Delta \mu > \Delta \mu_{\mathrm{thresh}}$, corresponding to a "switch-like" response of efficiency to increasing $\Delta \mu$. This prediction is borne out by recent \textit{in vitro} single-molecule measurements of dynein molecular motor step-size, which increase in a switch-like manner at a threshold ATP concentration \cite{dynein}, an observation that is otherwise difficult to explain (open circles in Fig. 2B). 
\par
Another difference between the heat engine efficiency and the dissipative chemical engine efficiency Eq.3 is that the latter is lower-bounded by $R_{\mathrm{L}}/(R_{\mathrm{S}}+R_{\mathrm{L}})$ in the near-equilibrium limit, whereas the heat engine efficiency tends to zero near equilibrium. Consider the special case in which a chemical engine is weakly driven: $\Delta \mu << kT$. In this case, the load dissipation simplifies to $\sigma_{\mathrm{L}} ={{R_{\mathrm{ L}}}\over{(R_{\mathrm{S}}+R_{\mathrm{L}})^2}}{\Delta \mu}^2  \leq {{1}\over{4R_{\mathrm{S}}}} {\Delta \mu}^2$, where the equality is achieved at the maximum load dissipation condition in which the load resistance is matched to that of the source: $R_{\mathrm{L,max}}=R_{\mathrm{S}}$. This condition is analogous to the maximum power transfer theorem of electronic circuit theory \cite{maxpowertransfer}. In this special case, Eq. 3 recovers the previously known result that $\eta_{\mathrm{chem}} \approx 1/2$ at maximum power in the near-equilibrium limit \cite{eff_increase,broeck}. Living systems operate at the other extreme far from equilibrium. For example, in the cell ATP concentration is maintained over ten orders of magnitude higher than its equilibrium value: $\Delta \mu \approx 24kT$, although values vary somewhat depending on cellular region and organism \cite{atp, atp2}. Eq.3 shows that living systems are intrinsically capable of being efficient over a wide range of load resistances corresponding to diverse functions (Fig. 2B) as long as $\Delta \mu/kT$ is sufficiently large. 
\par
Eq. 3 suggests that efficiency can be increased without increasing $\Delta \mu$ by increasing the load resistance (Fig. 2B). However, because increasing the resistance decreases the current, high efficiency near equilibrium (i.e. small $\Delta \mu$) would be expected to come at the cost of low throughput and power \cite{effberg}. Consider the processivity, a measure of time efficiency, which is the net flux (forward minus backward flux) divided by the forward flux: $P \equiv {{I}\over{f}}$. The tradeoff between energy efficiency $\eta_{\mathrm{chem}}$ and time efficiency $P$ can be quantified using Eq. 3 (See Supp. Materials):
 
 \begin{equation}
P \leq 1-e^{-{{\Delta \mu}\over {kT}} (1-\eta_{\mathrm{chem}})}.
 \end{equation} 
 This tradeoff is reminiscent of previous work on specific processes such as glycolysis \cite{doyle} and gene regulatory networks \cite{paulsson}, in which suppressing fluctuations comes at the cost of some measure of efficiency. 
 For a given $\Delta \mu$, the maximum $\eta_{\mathrm{chem}}$ and $P$ lie on a Pareto front (Fig.3). Higher $P$ must come at the expense of lower $\eta_{\mathrm{chem}}$, a tradeoff that is supported by \textit{in vivo} measurements of molecular motors \cite{myosin1,myosin2}. The new insight comes from considering Eq. 5 (and thus Eq. 3) in the far-from-equilibrium regime: the Pareto front sharpens as $\Delta \mu$ increases, such that for sufficiently strong driving there is effectively no longer a tradeoff. Within a Pareto front, an optimal tradeoff between energy efficiency and time efficiency can be defined by maximizing their sum: $\eta_{\mathrm{chem}}+P$ (solid black circles in Fig. 3; see Supp. Materials). Interestingly, a system with $R_{\mathrm{L}}=R_{\mathrm{S}}$ maximizes load dissipation and also coincide with maximum total efficiency near equilibrium; however, such a system sacrifices too much processivity for little energy efficiency gain far from equilibrium (open circles in Fig. 3), an example of a design principle that does not extrapolate from the weakly to the strongly driven regime.
\par
Another seemingly plausible mechanism for achieving high efficiency with small $\Delta \mu$ is to couple multiple driven transitions in series to achieve a larger effective total chemical potential difference. However, the generalized form of Eq. 3 for multiple sequential batteries shows that the efficiency scales inversely with the number of batteries, although the result could be more complex for arbitrary arrangements of batteries  (See Supp. Information). In particular, if $n$ identical driven transitions are connected in series (e.g. sequential hydrolysis of multiple ATP molecules), the efficiency in the processive regime is at most $1/n$. Therefore, efficient and processive dissipative machines require large units of energy "currency" $\Delta \mu$.
\newline
\textbf{Discussion.} Eq. 3 is a tight general bound relating the efficiency of dissipative machines to independent design parameters. These properties establish this equation as a chemical engine analog of Carnot's heat engine relation, and set it apart from previous work \cite{seifert_2011, max_eff_universal}. In contrast to Carnot's relation, in which the only property of the undriven engine is the ambient temperature, Eq. 3 includes information about the engine's design via the load resistance $R_{\mathrm{L}}$. This collective variable can be computed from knowing the detailed rate constants between the states of the engine, but it can also be experimentally measured as the linear response coefficient (i.e. susceptibility) of the net engine speed to $\Delta \mu$. By measuring the speed in the presence and absence of the desired load, $R_{\mathrm{S}}$ can also be determined. More broadly, Eq. 3 has implications for the design and evolution of molecular machines. If $\Delta \mu > \Delta \mu_{\mathrm{thresh}}$, high efficiency becomes a generic feature that does not require fine-tuning. Below the threshold, all molecular processes are inefficient. This threshold may therefore serve as a unique thermodynamic criterion for life. Because many biological processes are powered by the same sources of chemical energy, achieving large $\Delta \mu$ may have been an evolutionary transition step that "switched on" efficient energy transduction to all biomolecular processes of comparable load resistance.   
\newline

\noindent \textbf{Acknowledgements.} The author would like to thank Michael Trenfield for fruitful discussions regarding the theory and its implications, as well as Rama Ranganathan, Elliott Ross, Prashant Mishra, and Gaudenz Danuser for detailed feedback on the manuscript. This work was supported by the Cecil H. and Ida Green Foundation and the anonymous donor-supported UTSW High Risk/High Impact grant.

\bibliographystyle{unsrtnat}
\bibliography{Nonequil}

\newpage
\section{supplementary materials}
\subsection{Total dissipation rate}
The total rate of dissipation at steady state is:
\begin{equation}
    \sigma T = kT\sum_{ij} I_{ij} \ln {\Bigg[ {{P_i(k_{ij}+\alpha_{ij})}\over{P_j k_{ji}}} \Bigg]}
\end{equation}
I first recast this relation into the suggestive form: 

\begin{equation}
\sigma/k=\sum_{ij} I_{ij} (\ln[P_i/P_i^*] - \ln[P_j/P_j^*]) + \sum_{ij} I_{ij} \ln[1+\alpha_{ij}/k_{ij}],
\end{equation}
where $P_i^*$ and $P_j^*$ are the equilibrium probabilities of states $i$ and $j$ if all $\alpha$ (and therefore all $\Delta \mu$) are equal to zero; the ratio of these equilibrium probabilities are given by the ratio of the forward and backwards equilibrium rate constants between these states. Each term in the first summation is the current between adjacent states multiplied by the difference between a function evaluated at those states: in our circuit framework of symmetric resistors, Tellegen's Theorem dictates that such a sum equals zero at steady state for any circuit (Ref. 25 in Main Text). The nonzero sum is thus the steady-state dissipation rate (Eq. 2 in the main text):
\begin{equation}
 \sigma T= \sum_{ij} I_{ij} \Delta\mu_{ij},
 \end{equation} 
where the chemical potential difference $\Delta \mu_{ij}=kT\ln {\big[1+ {{\alpha_{ij}}\over{k_{ij}}} \big]}$.
\subsection{Dissipation rate of the load}
\subsubsection{single driven transition}
From Eq. 2, the total dissipation rate for a chemical engine with a single driven transition from state 1 to 2 (which we label as the source 'S') is:
\begin{equation}
 \sigma T= kT I_{\mathrm{S}} \ln {\Bigg[1+ {{\alpha_{\mathrm{S}}}\over{k_{\mathrm{S}}}} \Bigg]},
 \end{equation} 
where the current $I_{\mathrm{S}}$ flows through the source transition. The rate of dissipation of the driven transition (i.e. power dissipation in the battery) is:
\begin{equation}
 \sigma_{\mathrm{S}}T= kTI_{\mathrm{S}} \ln {\Bigg[{{(\alpha_{\mathrm{S}}+k_{\mathrm{S}})P_1}\over{P_2 k_{\mathrm{21}}}} \Bigg]},
 \end{equation}
 where the $k_{21}$ is the rate constant of the backwards reaction to $k_{\mathrm{S}}$ in the absence of driving (i.e. at equilibrium). Therefore, the rate of dissipation of the load is:
 \begin{equation}
 \sigma_{\mathrm{L}}T= \sigma T -\sigma_{\mathrm{S}}T=kT I_{\mathrm{S}} \ln{\Bigg[{{P_2 k_{\mathrm{21}}}\over{P_1 k_{\mathrm{S}}}}\Bigg]} .
 \end{equation} 
 
multiplying the numerator and denominator inside the logarithm by $e^{\beta G_1}$ and noting that $k_{\mathrm{S}}e^{\beta G_2}=k_{21}e^{\beta G_1}$, we obtain:
\begin{equation}
 \sigma_{\mathrm{L}}T=kT I_{\mathrm{S}}  \ln{\Bigg[{{P_2 e^{\beta G_2}}\over{P_1 e^{\beta G_1}}}\Bigg]} .
 \end{equation}
  Applying Ohm's law across the engine's cycle must yield zero change in the probability potential:
 \begin{equation}
 {{\alpha_{\mathrm{S}}}\over{k_{\mathrm{S}}}}P_1 e^{\beta G_1}-I_{\mathrm{S}}(R_{\mathrm{S}}+R_{\mathrm{L}})=0.
 \end{equation}
 Thus,
 \begin{equation}
 I_{\mathrm{S}}= {{\alpha_{\mathrm{S}}}\over{k_{\mathrm{S}}}}(R_{\mathrm{S}}+R_{\mathrm{L}})^{-1}P_1 e^{\beta G_1}.
 \end{equation}
 Now, applying Ohm's law across the load resistance:
 \begin{equation}
 P_2 e^{\beta G_2}-P_1 e^{\beta G_1}=I_{\mathrm{S}}R_{\mathrm{L}}.
 \end{equation}

Combining these two equations to solve for $P_2$:
 \begin{equation}
 P_2 e^{\beta G_2}= P_1 e^{\beta G_1}+{{\alpha_{\mathrm{S}}}\over{k_{\mathrm{S}}}}R_{\mathrm{L}}(R_{\mathrm{S}}+R_{\mathrm{L}})^{-1}P_1 e^{\beta G_1}.
 \end{equation}
 Therefore,
  \begin{equation}
 {{P_2 e^{\beta G_2}}\over{P_1 e^{\beta G_1}}}= {{R_{\mathrm{S}}+R_{\mathrm{L}}(1+{{\alpha_{\mathrm{S}}}\over{k_{\mathrm{S}}}})}\over{R_{\mathrm{S}}+R_{\mathrm{L}}}}.
 \end{equation}
 
Substituting for the probability ratio in the expression for the dissipation rate of the load gives:
\begin{equation}
 \sigma_{\mathrm{L}}T=kT I_{\mathrm{S}} \ln{\Bigg[{{R_{\mathrm{S}}+R_{\mathrm{L}}(1+{{\alpha_{\mathrm{S}}}\over{k_{\mathrm{S}}}})}\over{R_{\mathrm{S}}+R_{\mathrm{L}}}}\Bigg]}.
 \end{equation}
 Substituting for the chemical potential difference yields the dissipation rate of the load driven by a single battery given in the main text:
 \begin{equation}
 \sigma_{\mathrm{L}}T= kTI_{\mathrm{S}} \ln{\bigg[{{R_{\mathrm{S}}+R_{\mathrm{L}}e^{\Delta \mu/kT}}\over{R_{\mathrm{S}}+R_{\mathrm{L}}}}  \bigg]}.
\end{equation}

\subsubsection{multiple driven transitions in series}
Consider the more general case of $n$ driven transitions (batteries) in series. Denote the chemical potential difference and resistance of the $j$th battery by $\Delta \mu_j$ and $R_j$, respectively. $\Delta \mu_j$ drives the transition from state $j$ to state $j+1$. denote the forward and backward rate constants for transition $j$, in the absence of driving, by $k_j$ and ${\tilde{k}}_{j}$, respectively.   
The total rate of dissipation is:
\begin{equation}
 \sigma T= kT I_{\mathrm{S}} \sum_{j=1}^n\ln {\Bigg[1+ {{\alpha_{j}}\over{k_{j}}} \Bigg]}.
 \end{equation} 
The dissipation rate of the driven transitions (i.e. power dissipated in the batteries) is:
\begin{equation}
 \sigma_{\mathrm{S}}T= kT I_{\mathrm{S}} \sum_{j=1}^n\ln {\Bigg[{{(\alpha_{j}+k_{j})P_j}\over{P_{j+1} {\tilde{k}}_j}} \Bigg]}=kT I_{\mathrm{S}} \sum_{j=1}^n\ln {\Bigg[{{1+{{\alpha_{j}}\over{k_{j}}}\Bigg]}+kT I_{\mathrm{S}} \sum_{j=1}^n \ln {\Bigg[{{P_j}\over{P_{j+1}}} {{k_j}\over{\tilde{k}_j}}} \Bigg]}}.
 \end{equation}
 The last summation can be simplified by noting:
 
 \begin{equation}
 I_{\mathrm{S}} \sum_{j=1}^n \ln {\Bigg[{{P_j}\over{P_{j+1}}} {{k_j}\over{\tilde{k}_j}} \Bigg]} = I_{\mathrm{S}} \sum_{j=1}^n \ln {\Bigg[{{P_j}\over{P_{j+1}}} {{P_{j+1}^*}\over{P_{j}^*}} \Bigg]}=I_{\mathrm{S}}\ln {\Bigg[{{P_1}\over{P_{1}^*}} {{P_{n+1}^*}\over{P_{n+1}}} \Bigg]},
 \end{equation}
 where the first equality follows because the ratio of the undriven forward and backward rate constants gives the ratio of the corresponding probabilities of the states at equilibrium, and the second equality follows because all terms in the summation cancel except for the variables corresponding to the first and last states.
 Therefore, the rate of dissipation in the load is:
 \begin{equation}
 \sigma_{\mathrm{L}}T= \sigma T -\sigma_{\mathrm{S}}T= -kT I_{\mathrm{S}}\ln {\Bigg[{{P_1}\over{P_{1}^*}} {{P_{n+1}^*}\over{P_{n+1}}} \Bigg]}.
 \end{equation} 
 
  Applying Ohm's law across the $j$th battery:
 \begin{equation}
 {{\alpha_{j}}\over{k_{j}}}P_j e^{\beta G_j}-I_{\mathrm{S}}R_{j}=P_{j+1} e^{\beta G_{j+1}}-P_j e^{\beta G_j}.
 \end{equation}
 Thus,
 \begin{equation}
 \Big(1+{{\alpha_{j}}\over{k_{j}}}\Big)P_j e^{\beta G_j}-P_{j+1} e^{\beta G_{j+1}}=I_{\mathrm{S}}R_{j}.
 \end{equation}
 Multiply both sides by the product $\prod_{i=j+1}^{n}\Big(1+{{\alpha_{i}}\over{k_{i}}}\Big)$, and summing over all the batteries:
 \begin{equation}
 \sum_{j=1}^{n}\prod_{i=j+1}^{n}\Big(1+{{\alpha_{i}}\over{k_{i}}}\Big)\Big[\Big(1+{{\alpha_{j}}\over{k_{j}}}\Big)P_j e^{\beta G_j}-P_{j+1} e^{\beta G_{j+1}}\Big]=I_{\mathrm{S}}\sum_{j=1}^{n}\prod_{i=j+1}^{n}\Big(1+{{\alpha_{i}}\over{k_{i}}}\Big)R_{j}.
 \end{equation}
 
 All terms in the left sum cancel except for those corresponding to the first and last states:
 \begin{equation}
 \prod_{i=1}^{n}\Big(1+{{\alpha_{i}}\over{k_{i}}}\Big)P_1 e^{\beta G_1}-P_{n+1} e^{\beta G_{n+1}}=I_{\mathrm{S}}\sum_{j=1}^{n}\prod_{i=j+1}^{n}\Big(1+{{\alpha_{i}}\over{k_{i}}}\Big)R_{j}.
 \end{equation}
 
 Now, applying Ohm's law across the load resistance gives the current as a function of $P_1$ and $P_{n+1}$:
 \begin{equation}
 P_{n+1} e^{\beta G_{n+1}}-P_1 e^{\beta G_1}=I_{\mathrm{S}}R_{\mathrm{L}}.
 \end{equation}
Substituting for the current in the expression obtained from summing the batteries:

\begin{equation}
 \prod_{i=1}^{n}\Big(1+{{\alpha_{i}}\over{k_{i}}}\Big)P_1 e^{\beta G_1}-P_{n+1} e^{\beta G_{n+1}}=R_{\mathrm{L}}^{-1}\Big(P_{n+1} e^{\beta G_{n+1}}-P_1 e^{\beta G_1}\Big)\sum_{j=1}^{n}\prod_{i=j+1}^{n}\Big(1+{{\alpha_{i}}\over{k_{i}}}\Big)R_{j}.
 \end{equation}

Dividing both sides by $P_1e^{\beta G_1}$ and collecting terms:

\begin{equation}
 {{P_{n+1}}\over{P_{n+1}^*}}{{P_1^*}\over{P_{1}}}\Bigg(1+ R_{\mathrm{L}}^{-1}\sum_{j=1}^{n}\prod_{i=j+1}^{n}\Big(1+{{\alpha_{i}}\over{k_{i}}}\Big)R_{j}\Bigg)=\prod_{i=1}^{n}\Big(1+{{\alpha_{i}}\over{k_{i}}}\Big)+R_{\mathrm{L}}^{-1}\sum_{j=1}^{n}\prod_{i=j+1}^{n}\Big(1+{{\alpha_{i}}\over{k_{i}}}\Big)R_{j},
 \end{equation}
where the equilibrium probability ratios $P_{1}^*/P_{n+1}^* = e^{\beta(G_{n+1}-G_1)}$. Rearranging this expression, and substituting in the definition of the chemical potential difference:
\begin{equation}
 {{P_{n+1}}\over{P_{n+1}^*}}{{P_1^*}\over{P_{1}}}=\Bigg(1+ R_{\mathrm{L}}^{-1}\sum_{j=1}^{n}e^{\beta\sum_{i=j+1}^{n}\Delta \mu_i}R_{j}\Bigg)^{-1} \Bigg(e^{\beta\sum_{i=1}^{n}\Delta \mu_i}+R_{\mathrm{L}}^{-1}\sum_{j=1}^{n}e^{\beta\sum_{i=j+1}^{n}\Delta \mu_i}R_{j}\Bigg).
 \end{equation}
 
Substituting this probability ratio in the expression for the dissipation rate of the load for $n$ batteries in series:
\begin{equation}
 \sigma_{\mathrm{L}}T=kT I_{\mathrm{S}} \ln{\Bigg[{{e^{\beta\sum_{i=1}^{n}\Delta \mu_i}+R_{\mathrm{L}}^{-1}\sum_{j=1}^{n}e^{\beta\sum_{i=j+1}^{n}\Delta \mu_i}R_{j}}\over{1+R_{\mathrm{L}}^{-1}\sum_{j=1}^{n}e^{\beta\sum_{i=j+1}^{n}\Delta \mu_i}R_{j}}} \Bigg]}.
 \end{equation}
 In the special case in which all the driven transitions are identical, $\Delta \mu_j = \Delta \mu$ and $R_j = R_{\mathrm{S}}$. Then the entropy production rate simplifies to:
 \begin{equation}
 \sigma_{\mathrm{L}}T=kT I_{\mathrm{S}} \ln{\Bigg[{{R_{\mathrm{S}}{{e^{n\beta\Delta \mu}-1}\over{e^{\beta\Delta \mu}-1}}+e^{n\beta\Delta \mu}R_{\mathrm{L}}}\over{R_{\mathrm{S}}{{e^{n\beta\Delta \mu}-1}\over{e^{\beta\Delta \mu}-1}}+R_{\mathrm{L}}}} \Bigg]}.
 \end{equation}
 As expected, this expression reduces to the single battery dissipation rate derived in the previous section in the case that $n=1$.
 \par
 The maximum efficiency is given by
 \begin{equation}
      \eta_{\mathrm{chem}}\leq {{\sigma_{\mathrm{L}}}\over{\sigma}}={{kT}\over{n\Delta \mu}} \ln{\Bigg[{{R_{\mathrm{S}}{{e^{n\beta\Delta \mu}-1}\over{e^{\beta\Delta \mu}-1}}+e^{n\beta\Delta \mu}R_{\mathrm{L}}}\over{R_{\mathrm{S}}{{e^{n\beta\Delta \mu}-1}\over{e^{\beta\Delta \mu}-1}}+R_{\mathrm{L}}}} \Bigg]}
 \end{equation}
In the processive regime ($R_{\mathrm{S}}>R_{\mathrm{L}}$), this simplifies to 
\begin{equation}
      \eta_{\mathrm{chem}} < {{kT}\over{n\Delta \mu}} \ln{\Bigg[{{R_{\mathrm{L}}}\over{R_{\mathrm{S}}}}e^{\beta\Delta \mu}\Bigg]} < {{1}\over{n}}.
\end{equation}
Therefore, as stated in the main text, the maximum efficiency of a chemical engine powered by multiple driven transitions in series is the reciprocal of the number of driven transitions.

\subsection{Inflection point of the efficiency function}
Define reduced variables: $u \equiv \Delta \mu/kT$ and $r \equiv R_{\mathrm{S}}/R_{\mathrm{L}}$. In these reduced variables, the maximum efficiency(Eq. 3 in Main Text) is:
\begin{equation}
    \eta_{\mathrm{chem, max}} = {{1}\over{u}}\ln{\bigg[ {{e^{u}+r}\over{1+r}}\bigg]}.
\end{equation}
Differentiating twice with respect to $u$ yields the curvature of the efficiency function:
\begin{equation}
{{\partial^2 \eta_{\mathrm{chem, max}}}\over{\partial u^2}} ={{2\ln{\big[{{e^u+r}\over{1+r}} \big]}}\over{u^3}}+ {{e^u}\over{(e^u+r)u}}-{2{e^u}\over{(e^u+r)u^2}}-{{e^{2u}}\over{(e^u+r)^2 u}}
\end{equation}

For $\eta_{\mathrm{chem, max}}$ to have an inflection point, there must be a value of $u$ for which the curvature is zero. In the limit of vanishing $u$, such that $e^u$ approaches $1+u$, the curvature is: 
\begin{equation}
{{\partial^2 \eta_{\mathrm{chem, max}}}\over{\partial u^2}} = {{r(r-1)}\over{3(r+1)^3}} + O(u)
\end{equation}
Therefore, if $r>1$ (i.e. $R_{\mathrm{S}} > R_{\mathrm{L}}$), the curvature is positive in the limit of very small $u$ (i.e. $\Delta \mu << kT$).
Now consider the other limit of large $u >> 1$. In this case, define the variable $x \equiv r/e^u$. Then the curvature can be rewritten:

\begin{equation}
{{\partial^2 \eta_{\mathrm{chem, max}}}\over{\partial u^2}} ={{2(u+\ln{\big[{{1+x}\over{1+r}} \big]})}\over{u^3}}+ {{1}\over{(1+x)u}}-{{2}\over{(1+x)u^2}}-{{1}\over{(1+x)^2 u}}
\end{equation}
In the limit of large $u$, $x$ will approach zero, and the leading order curvature is:
\begin{equation}
{{\partial^2 \eta_{\mathrm{chem, max}}}\over{\partial u^2}} = {{2}\over{u^3}}\ln{\bigg[{{1}\over{1+r}} \bigg]} + O(x),
\end{equation}
which approaches zero from below because $r$ is positive. Therefore, if $r > 1$, the efficiency must pass through an inflection point because the curvature is positive for small $\Delta \mu$ and negative for large $\Delta \mu$.
\par
To find the inflection point, make the guess that it occurs near the condition $u=\Delta \mu /kT \approx \ln{[r]}$, such that the logarithm in the first term of the curvature equation is of order unity or smaller. In that case, the first term scales with the inverse cube of $u$; in the limit of large $u$, we can ignore this term to obtain the inflection point condition:
\begin{equation}
{{\partial^2 \eta_{\mathrm{chem, max}}}\over{\partial u^2}} \Bigg|_{u=u_{\mathrm{thresh}}} \approx {{e^{u}}\over{(e^{u}+r){u}}}-{2{e^u}\over{(e^u+r)u^2}}-{{e^{2u}}\over{(e^u+r)^2 u}} = 0,
\end{equation}
Which yields the solution:

\begin{equation}
u_{\mathrm{thresh}} \approx 2-\textit{W}_{-1}(-2e^2/r),
\end{equation}
where $\textit{W}_{-1}$ is the lower real branch of the Lambert W function. In the limit of large $r$, this is approximated by:
\begin{equation}
u_{\mathrm{thresh}} \approx \ln{[r]}+\ln{\bigg[{{1}\over{2}}\ln{\big[{{r}\over{2}}\big]}-1\bigg]}.
\end{equation}
This result is self-consistent with the ansatz at the beginning of this section that requires $u \approx \ln{[r]}$.
To leading order, the inflection point is give by Eq. 4 in the main text:
\begin{equation}
\Delta \mu_{\mathrm{thresh}} \approx kT\ln{\bigg[{{R_{\mathrm{S}}}\over{R_{\mathrm{L}}}}\bigg]}.
\end{equation}

\subsection{Tradeoff between efficiency and processivity}
Starting with the efficiency relation (Eq. 3 in the main text):
\begin{equation}
\eta_{\mathrm{chem}} \leq {{kT}\over{\Delta \mu}} \ln{\Bigg[{{{{R_{\mathrm{S}}}\over{R_{\mathrm{L}}}}+e^{{{\Delta \mu} \over{kT}}}}\over{{{R_{\mathrm{S}}}\over{R_{\mathrm{L}}}}+1}}  \Bigg]}.
\end{equation}
Rearrange to obtain:
\begin{equation}
-{{\Delta \mu}\over{kT}}(1-\eta_{\mathrm{chem}}) \leq  \ln{\Bigg[{{{{R_{\mathrm{S}}}\over{R_{\mathrm{L}}}}+e^{{{\Delta \mu} \over{kT}}}}\over{{{R_{\mathrm{S}}}\over{R_{\mathrm{L}}}}+1}}  \Bigg]}-{{\Delta \mu}\over{kT}}=\ln{\Bigg[{{R_{\mathrm{S}}e^{-{{\Delta \mu}\over{kT}}}+R_{\mathrm{L}}}\over{R_{\mathrm{S}}+R_{\mathrm{L}}}}\Bigg]}.
\end{equation}
exponentiating both sides, subtracting from one, and simplifying:
\begin{equation}
1-e^{-{{\Delta \mu}\over{kT}}(1-\eta_{\mathrm{chem}})} \geq  1-{{R_{\mathrm{S}}e^{-{{\Delta \mu}\over{kT}}}+R_{\mathrm{L}}}\over{R_{\mathrm{S}}+R_{\mathrm{L}}}} = {{R_{\mathrm{S}}(e^{{{\Delta \mu}\over{kT}}}-1)}\over{e^{{{\Delta \mu}\over{kT}}}(R_{\mathrm{S}}+R_{\mathrm{L}})}}.
\end{equation}
Substituting for the chemical potential difference:
\begin{equation}
1-e^{-{{\Delta \mu}\over{kT}}(1-\eta_{\mathrm{chem}})} \geq {{R_{\mathrm{S}}({{\alpha_{\mathrm{S}}}\over{k_{\mathrm{S}}}})}\over{(1+{{\alpha_{\mathrm{S}}}\over{k_{\mathrm{S}}}})(R_{\mathrm{S}}+R_{\mathrm{L}})}}={{{\alpha_{\mathrm{S}}}\over{k_{\mathrm{S}}}}\over{(R_{\mathrm{S}}+R_{\mathrm{L}})}} {{e^{\beta G_1}}\over{k_{\mathrm{S}}+\alpha_{\mathrm{S}}}},
\end{equation}
where we substituted for $R_{\mathrm{S}}=e^{\beta G_1}/k_{\mathrm{S}}$. Ohm's law taken across the engine cycle gives:
\begin{equation}
{{\alpha_{\mathrm{S}}}\over{k_{\mathrm{S}}}}P_1e^{\beta G_1}=I_{\mathrm{S}}(R_{\mathrm{S}}+R_{\mathrm{L}}),
\end{equation}
which gives the relation:
\begin{equation}
{{{\alpha_{\mathrm{S}}}\over{k_{\mathrm{S}}}}\over{(R_{\mathrm{S}}+R_{\mathrm{L}})}}={{I_{\mathrm{S}}}\over{P_1e^{\beta G_1}}}.
\end{equation}
Substituting this relation into the inequality:
\begin{equation}
1-e^{-{{\Delta \mu}\over{kT}}(1-\eta_{\mathrm{chem}})} \geq {{I_{\mathrm{S}}}\over{P_1e^{\beta G_1}}}{{e^{\beta G_1}}\over{k_{\mathrm{S}}+\alpha_{\mathrm{S}}}} = {{I_{\mathrm{S}}}\over{P_1(k_{\mathrm{S}}+\alpha_{\mathrm{S}})}}.
\end{equation}
Noting that the denominator in the right hand side is the total forward flux $f_{\mathrm{S}}=P_1(k_{\mathrm{S}}+\alpha_{\mathrm{S}})$. The net flux $I_{\mathrm{S}}$ divided by the total forward flux $f_{\mathrm{S}}$ is the processivity $P$. Therefore, we obtain Eq. 5 in the main text:
 \begin{equation}
P \leq 1-e^{-{{\Delta \mu}\over {kT}} (1-\eta_{\mathrm{chem}})}.
 \end{equation}

\subsection{Dynein efficiency}
The maximum \textit{in vitro} efficiency of dynein under different ATP concentrations was obtained from Ref. 28. Dynein step size at different concentrations was obtained from Fig. 3E in Ref. 28. The step size was converted to efficiency by multiplying the step size by the reported stall force of 1.85 pN, and dividing by the corresponding $\Delta \mu$. $\Delta \mu$ was estimated from the \textit{in vivo} value, with an [ATP]-dependent correction $\ln{[\mathrm{[ATP]/[ATP]}_{vivo}]}$ with respect to  an \textit{in vivo} ATP concentration of 10mM.

\end{document}